\documentclass[aps,prb,reprint,amsfonts,amsmath,amssymb,longbibliography,superscriptaddress]{revtex4-2}
\usepackage{hyperref}
\usepackage{graphicx}
\usepackage{bm}
\usepackage{newtxtext}
\usepackage[varg]{newtxmath}
\usepackage[normalem]{ulem}
\usepackage{braket}
\hypersetup{colorlinks=true,linkcolor=blue,citecolor=blue,urlcolor=blue}
\usepackage{mathtools}
\usepackage{mleftright}
\mleftright

\DeclareMathOperator{\diag}{diag}
\DeclareMathOperator*{\Tr}{Tr}
\DeclareMathOperator{\sgn}{sgn}

\begin{document}

\title{High harmonic generation from electrons moving in topological spin textures}
\author{Atsushi Ono}
\affiliation{\mbox{Department of Physics, Tohoku University, Sendai 980-8578, Japan}}
\author{Shun Okumura}
\affiliation{\mbox{Department of Applied Physics, The University of Tokyo, Hongo, Tokyo 113-8656, Japan}}
\author{Shohei Imai}
\affiliation{\mbox{Department of Physics, Graduate School of Science, The University of Tokyo, 7-3-1 Hongo, Tokyo 113-0033, Japan}}
\author{Yutaka Akagi}
\affiliation{\mbox{Department of Physics, Graduate School of Science, The University of Tokyo, 7-3-1 Hongo, Tokyo 113-0033, Japan}}
\date{\today}

\begin{abstract}
High harmonic generation (HHG) is a striking phenomenon, which reflects the ultrafast dynamics of electrons.
Recently, it has been demonstrated that HHG can be used to reconstruct not only the energy band structure but also the geometric structure characterized by the Berry curvature.
Here, we numerically investigate HHG arising from electrons coupled with a topological spin texture in a spin scalar chiral state where time reversal symmetry is broken.
In this system, a sign change in scalar chirality alters the sign of the Berry curvature while keeping the energy band structure unchanged, allowing us to discuss purely geometrical effects on HHG.
Notably, we found that, when the optical frequency is significantly lower than the energy gap, the sign of scalar chirality largely affects the longitudinal response parallel to the optical field rather than the transverse response.
Our analysis suggests that this can be attributed to interband currents induced by the recombination of electron--hole pairs whose real-space trajectories are modulated by the anomalous velocity term.
\end{abstract}

\maketitle

\section{Introduction}
With the advancement of laser technology, ultrafast phenomena in the subfemtosecond and attosecond domains are being actively researched \cite{Agostini2004, Gallmann2012, Krausz2009, Kruchinin2018, Ghimire2019, Amini2019, Li2023, Borsch2023, Na2023, DelaTorre2021, Filippetto2022, Boschini2024, Oka2018, Murakami2023c}.
High harmonic generation (HHG) and high-order sideband generation (HSG) are representative examples, and recent research has progressed in solids such as semiconductors \cite{Ghimire2011, Schubert2014, Hohenleutner2015, Garg2016, Langer2016, Langer2017, Yoshikawa2017, Mrudul2020, Xia2021, Tamaya2016, Sato2021a, Murakami2022, Murakami2023, Floss2018, Sekiguchi2023, Kono1997, Zaks2012, Langer2018, Uchida2018a, Borsch2020, Nagai2020a, Costello2021, Uchida2024}, strongly correlated electron systems \cite{Silva2018a, Murakami2018a, Imai2020a, Tancogne-Dejean2018, Granas2022, Nag2019a, Murakami2021, Orthodoxou2021, Lysne2020a, Bionta2021, Shao2022, Udono2022, Hansen2022a, Hansen2022b, Uchida2022, Murakami2022b, Kofuji2023, Alcala2022, Zhu2021, Fauseweh2020, Nakano2023, Lange2024}, and magnetic materials \cite{Zhang2018m, Takayoshi2019, Ikeda2019, Pattanayak2022, Kanega2021, Allafi2024, Ly2023, Ly2022, Varela-Manjarres2023, Okumura2021}.
In HHG and HSG, high-order harmonics are literally generated, and the details of their spectra and chirping have been well captured by the three-step model \cite{Corkum1993, Lewenstein1994, Vampa2015a, Nayak2019a, Imai2022}.
According to this model, the elementary processes of HHG, for example, consist of (i) ionization of electrons to the vacuum or excitation to the conduction bands, (ii) acceleration, and (iii) recombination of the electrons or the electron--hole pairs.
Therefore the electronic structure is embedded in the high harmonic spectrum, and using this property, all-optical reconstruction of energy bands through HHG and HSG has been proposed and experimentally demonstrated \cite{Luu2015, Vampa2015, Yu2018f, Li2020dm, Uzan-Narovlansky2022, Imai2022}.

Recently, it has become increasingly clear that HHG can be used to extract not only the energy band structure but also the geometric structure of electrons, characterized by the Berry curvature or the Berry phase \cite{Liu2016c, Schmid2021, Lv2021, Luu2018b, Bai2020a, Silva2019a, Lou2021, Banks2017, Yue2023, Uzan-Narovlansky2024, Wu2017}.
As is well known, the Berry curvature appears in systems where either spatial inversion symmetry or time reversal symmetry is broken.
Hitherto, high harmonics dependent on the Berry curvature have been observed in systems with broken spatial inversion symmetry, for example, in a monolayer $\mathrm{MoS_2}$ \cite{Liu2016c} and the surface states of a topological insulator $\mathrm{Bi_2Te_3}$ \cite{Schmid2021}.
Additionally, in a Weyl semimetal $\mathrm{WP_2}$ \cite{Lv2021}, the Berry curvature has been successfully reconstructed in reciprocal space.
However, studies on the effects of geometric structures in HHG have been scarce for systems with broken time reversal symmetry \cite{Medic2024}.

Systems exhibiting nonzero Berry curvature due to broken time reversal symmetry include those with what are called topological spin textures.
For example, in skyrmion crystals, interesting perturbative linear responses such as the topological Hall effect \cite{Nagaosa2013} and the magneto-optical effect \cite{Hayashi2021, Sorn2021, Feng2020k, Kato2023, Li2024} have been reported and discussed.
However, nonlinear responses, including HHG or HSG, have received limited attention \cite{Esin2022, Hori2024}, even though the magnetic structure is expected to be embedded in the high harmonic spectrum through the dynamics of electrons coupled with topological spin textures.

A spin scalar chiral state can be considered as one of the simplest topological spin textures.
It features a four-sublattice magnetic order [Figs.~\ref{fig:sketch}(a) and \ref{fig:sketch}(b)], where the Chern number of each energy band takes on an integer value, leading to the emergence of the anomalous (topological) Hall effect \cite{Martin2008, Akagi2010}.
Hence, this state can be viewed as a skyrmion crystal state with the smallest magnetic unit cell.
Last year, two groups experimentally reported that the scalar chiral state is realized in $\mathrm{CoTa_3S_6}$ and $\mathrm{CoNb_3S_6}$ \cite{Takagi2023, Park2023, Park2024}, attracting significant interest.
In this study, we numerically analyze HHG in the scalar chiral state.
We found that the transverse response, as naively expected to reflect the Berry curvature, indeed appears.
Furthermore, we discovered that the sign of the Berry curvature is reflected in the longitudinal response even in cases where it predominates over the transverse response.
This finding differs from the effects of anomalous velocity in intraband currents that have been discussed previously.
We argue that the anomalous velocity modulates the recombination conditions of electron--hole pairs, potentially changing the interband current spectrum, on the basis of an analysis of the real-space dynamics of electron--hole pairs.

The rest of this paper is organized as follows.
In Sec.~\ref{sec:method}, we introduce our model and methods, and in Sec.~\ref{sec:results}, we present numerical results.
Section~\ref{sec:equil} provides an overview of the equilibrium properties, focusing particularly on its geometric structure, and Sec.~\ref{sec:hhg} displays the high harmonic spectrum obtained from real-time evolution.
The analysis of the real-space dynamics of electron--hole pairs is conducted in Sec.~\ref{sec:trajectory}.
Section~\ref{sec:material} discusses HHG for parameters close to those for $\mathrm{CoTa_3S_6}$ and $\mathrm{CoNb_3S_6}$.
Sections~\ref{sec:discussion} and \ref{sec:summary} are respectively devoted to the discussion and summary.
Appendices~\ref{sec:near_resonant}--\ref{sec:perturbative} provide results for near-resonant and circular polarization driving, as well as comparisons with the $120^{\mathrm{\circ}}$ N\'eel state, where the Berry curvature is absent, and with perturbative responses.

\section{Model and method} \label{sec:method}
To examine the dynamics of electrons coupled with spin textures, we consider the ferromagnetic Kondo lattice model on a two-dimensional triangular lattice.
The Hamiltonian is defined by
\begin{align}
\mathcal{H}
= \sum_{ijs} h_{ij} c_{is}^\dagger c_{js} - J_{\mathrm{K}} \sum_{iss'} \bm{S}_i \cdot \bm{\sigma}_{ss'} c_{is}^\dagger c_{is'},
\label{eq:fklm}
\end{align}
where $c_{is}^\dagger$ is a creation operator of an electron at site $i$ with spin $s$, $\bm{\sigma}$ is a three-component vector of the Pauli matrices, and $\bm{S}_i$ is a classical vector describing a localized spin at site $i$.
The transfer integral and the exchange interaction strength are denoted by $h_{ij}$ and $J_{\mathrm{K}}$, respectively.
Considering that the electron dynamics induced by external fields occur on time scales of the order of subpicoseconds, we assume in this study that a magnetic order of $\{\bm{S}_i\}$ is not disturbed by electron motions; that is, each $\bm{S}_i$ is frozen during the real-time evolution of the electrons.

\begin{figure}[t]\centering
\includegraphics[scale=1]{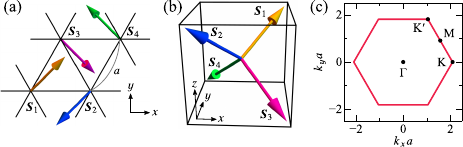}
\caption{(a)~Magnetic unit cell of the four-sublattice spin scalar chiral state and (b) configuration of the four localized spin vectors, adapted from Ref.\ \cite{Ono2023}.
(c)~Magnetic Brillouin zone.}
\label{fig:sketch}
\end{figure}

While the Hamiltonian in Eq.~\eqref{eq:fklm} is invariant under the global rotation of $\{\bm{S}_i\}$ and $\bm{\sigma}$, we explicitly define the four-sublattice scalar chiral state as
\begin{gather}
\bm{S}_1 = \frac{(1,1,1)}{\sqrt{3}}, \quad
\bm{S}_2 = \frac{(-1,-1,1)}{\sqrt{3}}, \notag \\
\bm{S}_3 = \frac{(1,-1,-1)}{\sqrt{3}}, \quad
\bm{S}_4 = \frac{(-1,1,-1)}{\sqrt{3}},
\label{eq:def_spindirections}
\end{gather}
where $\bm{S}_m$ represents the localized spin of sublattice $m$ instead of site $i$ (see Fig.~\ref{fig:sketch}).
For this choice of $\{\bm{S}_m\}$, the scalar chirality, defined by
\begin{align}
\chi = \bm{S}_1 \cdot (\bm{S}_2 \times \bm{S}_3) + \bm{S}_4 \cdot (\bm{S}_3 \times \bm{S}_2),
\end{align}
has a positive value of $\chi = 8/(3\sqrt{3}) \equiv \chi_{0}$.
The sign of $\chi$ is changed by time reversal: $\bm{S}_m \mapsto -\bm{S}_m$ for all $m$; this operation is equivalent to $J_{\mathrm{K}} \mapsto -J_{\mathrm{K}}$ in the electron system, while the band structure remains unchanged, since the localized spins $\bm{S}_m$ are treated classically.

Assuming sublattice structure, we can express the Hamiltonian in reciprocal space as
\begin{align}
\mathcal{H} = \sum_{\bm{k}} \sum_{ss'} \sum_{mm'} h_{sm,s'm'}(\bm{k}) c_{\bm{k}sm}^\dagger c_{\bm{k}s'm'},
\end{align}
where $\bm{k}$ denotes momentum, and the indices $s$ and $m$ correspond respectively to the spin and sublattice degrees of freedom.
In the four-sublattice scalar chiral state, each energy band is doubly degenerated in the whole magnetic Brillouin zone (BZ) [Fig.~\ref{fig:sketch}(c)], and $h(\bm{k})$ is an $8\times 8$ matrix that can be block diagonalized by the unitary transformation:
\begin{align}
U^\dagger h(\bm{k}) U
= h_+(\bm{k}) \oplus h_-(\bm{k}),
\label{eq:hamiltonian_block}
\end{align}
into two $4\times 4$ matrices $h_+$ and $h_-$.
Adopting $\{\bm{S}_m\}$ in Eq.~\eqref{eq:def_spindirections}, we can choose the unitary matrix $U$ as
\begin{align}
U &= \text{(spin)} \otimes \text{(sublattice)} \notag \\
&= \frac{1}{\sqrt{2}} {\begin{bmatrix}
-\mathrm{i} & 0 & 0 & 0 & \mathrm{i} & 0 & 0 & 0 \\
0 & -\mathrm{i} & 0 & 0 & 0 & \mathrm{i} & 0 & 0 \\
0 & 0 & -\mathrm{i} & 0 & 0 & 0 & \mathrm{i} & 0 \\
0 & 0 & 0 & -\mathrm{i} & 0 & 0 & 0 & \mathrm{i} \\
0 & 0 & 0 & 1 & 0 & 0 & 0 & 1 \\
0 & 0 & 1 & 0 & 0 & 0 & 1 & 0 \\
0 & 1 & 0 & 0 & 0 & 1 & 0 & 0 \\
1 & 0 & 0 & 0 & 1 & 0 & 0 & 0
\end{bmatrix}},
\end{align}
which diagonalizes another unitary matrix $V$ that commutes with $h(\bm{k})$.
The latter unitary matrix,
\begin{align}
V
= \sigma_y \otimes {\begin{bmatrix}
0 & 0 & 0 & 1 \\
0 & 0 & 1 & 0 \\
0 & 1 & 0 & 0 \\
1 & 0 & 0 & 0
\end{bmatrix}},
\end{align}
represents a mirror reflection of electron spins with respect to the $zx$ plane and a permutation of sublattice spins: $\bm{S}_1 \leftrightarrow \bm{S}_4$ and $\bm{S}_2 \leftrightarrow \bm{S}_3$, and it is diagonalized as
\begin{align}
U^\dagger V U = \diag(+1,+1,+1,+1,-1,-1,-1,-1).
\end{align}
The explicit form of $h_\pm$ is given by
\begin{align}
h_\pm(\bm{k}) &= -\frac{J_{\mathrm{K}}}{\sqrt{3}} \sigma_z \otimes \sigma_0
\pm \frac{J_{\mathrm{K}}}{\sqrt{3}} (\sigma_x+\sigma_y)\otimes \sigma_y \notag \\
&\quad - 2h_1\cos(k_x) \sigma_0 \otimes \sigma_x \notag \\
&\quad - 2h_1\cos\left(\frac{k_x+\sqrt{3}k_y}{2}\right) \sigma_x \otimes \sigma_0 \notag \\
&\quad - 2h_1\cos\left(\frac{k_x-\sqrt{3}k_y}{2}\right) \sigma_x \otimes \sigma_x ,
\end{align}
where $\sigma_0$ denotes the identity matrix.
Here and throughout the paper, we consider only the transfer integral between the nearest neighbor sites,  $h_{ij} = -h_1$.
Given the above, we can write the Hamiltonian in Eq.~\eqref{eq:fklm} as
\begin{align}
\mathcal{H} = \sum_{\bm{k}} \sum_{n=1}^{4} {\left[ \varepsilon_n^+(\bm{k}) a_{\bm{k}n}^\dagger a_{\bm{k}n} + \varepsilon_n^-(\bm{k}) b_{\bm{k}n}^\dagger b_{\bm{k}n} \right]}
\end{align}
with $\varepsilon_n^+(\bm{k}) = \varepsilon_n^-(\bm{k}) \equiv \varepsilon_n(\bm{k})$.
Here, $\varepsilon_n^\pm(\bm{k})$ is the $n$th eigenenergy of $h_\pm(\bm{k})$, and $a_{\bm{k}n}^\dagger$ and $b_{\bm{k}n}^\dagger$ are creation operators of electrons associated with $c_{\bm{k}sm}^\dagger$ through a unitary transformation.
The block diagonal form of $h(\bm{k})$ in Eq.~\eqref{eq:hamiltonian_block} facilitates the efficient computation of real-time dynamics and enables the calculation of the Berry curvature $\bm{B}_n = (0,0,B_n)$ through the following formula:
\begin{align}
B_n(\bm{k})
= \sum_{m(\neq n)} \frac{\langle \bm{k}n \vert v_x(\bm{k}) \vert \bm{k}m \rangle \langle \bm{k}m \vert v_y(\bm{k}) \vert \bm{k}n \rangle - \text{c.c.}}{[\varepsilon_n(\bm{k}) - \varepsilon_m(\bm{k})]^2}
\label{eq:berry}
\end{align}
with $\bm{v}(\bm{k}) = \partial_{\bm{k}} h_+(\bm{k})\ [= \partial_{\bm{k}} h_-(\bm{k})]$, where $\vert \bm{k}n \rangle$ is an energy eigenstate satisfying $h_+(\bm{k}) \vert\bm{k}n\rangle = \varepsilon_n(\bm{k}) \vert\bm{k}n\rangle$.
Note that $B_n(\bm{k})$ remains unchanged even if we adopt the eigenstates of $h_-(\bm{k})$, and the Chern number of the doubly degenerated $n$th band is given by
\begin{align}
\mathcal{C}_n = \int_{\text{BZ}} \frac{\mathrm{d}^2k}{2\pi}\, 2B_n(\bm{k}),
\label{eq:chern}
\end{align}
where BZ stands for the magnetic BZ depicted in Fig.~\ref{fig:sketch}(c).
We also introduce the linear optical conductivity \cite{Maekawa2004},
\begin{align}
\sigma_{\alpha\beta}(\omega)
&= \frac{\mathrm{i} T_{\alpha\beta} - \mathrm{i} \chi_{\alpha\beta}(\omega+\mathrm{i}\eta)}{\omega+\mathrm{i}\eta}, \label{eq:conductivity}
\end{align}
with $\eta$ being a positive infinitesimal, where
\begin{align}
T_{\alpha\beta}
&= \int_{\text{BZ}} \frac{\mathrm{d}^2k}{(2\pi)^2} \Tr[ \rho_{\bm{k}} \partial_{k_\alpha} \partial_{k_\beta} h(\bm{k}) ], \label{eq:stresstensor} \\
\chi_{\alpha\beta}(z)
&= 2 \int_{\text{BZ}} \frac{\mathrm{d}^2k}{(2\pi)^2} \sum_{mn} \frac{f_n(\bm{k})-f_m(\bm{k})}{\varepsilon_m(\bm{k}) - \varepsilon_n(\bm{k}) - z} \notag \\
&\quad \times \langle \bm{k}n \vert v_\alpha(\bm{k}) \vert \bm{k}m \rangle \langle \bm{k}m \vert v_\beta(\bm{k}) \vert \bm{k}n \rangle. \label{eq:kernel}
\end{align}
Here, $\rho_{\bm{k},mn} = \langle c_{\bm{k}n}^\dagger c_{\bm{k}m} \rangle$ represents a one-body density matrix of electrons, $f_n(\bm{k}) = \langle a_{\bm{k}n}^\dagger a_{\bm{k}n} \rangle = \langle b_{\bm{k}n}^\dagger b_{\bm{k}n} \rangle$ is the Fermi distribution function for the $n$th band, and $T_{\alpha\beta}$ in Eq.~\eqref{eq:stresstensor} is called a stress tensor.
The prefactor $2$ in Eq.~\eqref{eq:kernel} counts the equal contribution from the eigenstates of $h_-$.

The real-time dynamics are governed by the von Neumann equation with a phenomenological relaxation term,
\begin{align}
\frac{\mathrm{d}\rho_{\bm{k}}}{\mathrm{d}t}
= -\mathrm{i} [h(\bm{k}), \rho_{\bm{k}}] - \varGamma (\rho_{\bm{k}} - \rho_{0,\bm{k}}),
\label{eq:vonNeumann}
\end{align}
with $t$ representing time.
Here, $\rho_{0,\bm{k}}$ denotes the one-body density matrix in the ground state for a given $\bm{k}$, and $\varGamma$ represents the relaxation rate.
We assume that the initial state is the ground state, that is, $\rho_{\bm{k}}(-\infty) = \rho_{0,\bm{k}}$.
Given our focus on the dynamics of electrons driven by optical fields, we consider only the coupling between the electric field and the electrons.
The vector potential $\bm{A}(t)$ is introduced through the Peierls substitution: $\bm{k}(t) = \bm{k} - \bm{A}(t)$, and the electric field $\bm{F}$ is determined by $\bm{F}(t) = -\partial_t \bm{A}(t)$.
In this study, we apply a continuous wave described by the following vector potential:
\begin{align}
A_\alpha(t) &= -\frac{\bm{F}_{0,\alpha}}{\varOmega} \sin(\varOmega t - \phi_\alpha) \times
{\left\{ \begin{array}{cl}
\mathrm{e}^{-t^2/(2\tau^2)} & (t < 0) \\
1 & (t \geq 0)
\end{array} \right.}
\label{eq:cw}
\end{align}
for $\alpha = x, y$, where $\bm{F}_0 = (F_{0,x}, F_{0,y})$, $\varOmega$, $\phi_{\alpha}$, and $\tau$ represent the electric field amplitude, frequency, phase, and ramp time, respectively.
Linear polarization is characterized by
\begin{align}
F_{0,x} = F_0 \cos\psi, \quad
F_{0,y} = F_0 \sin\psi, \quad
\phi_{x} = \phi_{y} = \phi,
\label{eq:lp}
\end{align}
where $\psi$ denotes the polarization angle measured from the $x$ axis [see also Fig.~\ref{fig:sketch}(a)].
On the other hand, left/right circular polarization (LCP/RCP) is described by
\begin{align}
F_{0,x} = F_{0,y} = F_0, \quad
(\phi_x, \phi_y) = {\left\{ \begin{array}{cl}
(-\pi/2,0) & (\text{LCP}) \\
(+\pi/2,0) & (\text{RCP}).
\end{array} \right.}
\label{eq:cp}
\end{align}

The electric current density $\bm{J}(t)$ is defined by
\begin{align}
\bm{J}(t) = \frac{1}{N\mathcal{A}} \sum_{\bm{k}} \Tr[ \rho_{\bm{k}} \bm{v}(\bm{k}) ],
\end{align}
where $N$ stands for the number of $\bm{k}$-points.
The area of the magnetic unit cell is denoted by $\mathcal{A}$, and in the presence of four-sublattice orders, $\mathcal{A} = 2\sqrt{3} a^2$ with $a$ being the lattice constant.
The intensity of electromagnetic radiation is proportional to
\begin{align}
I(\omega) = I_x(\omega) + I_y(\omega), \quad
I_\alpha(\omega) = \omega^2 |J_\alpha(\omega)|^2
\end{align}
for $\alpha = x,y$, where $\bm{J}(\omega)$ is the Fourier transformation of $\bm{J}(t)$.

The von Neumann equation~\eqref{eq:vonNeumann} is numerically solved using the fourth-order Runge--Kutta method with a time step of $\delta t = 0.01 \hbar/h_1$.
The number of $\bm{k}$-points is set to $N = 50^2$ unless otherwise specified, for which we confirmed the convergence.
In this paper, the nearest neighbor transfer integral $h_1$, the Dirac constant $\hbar$, the electric charge $e$, and the lattice constant $a$ are set to one.
Energy, time, electric current density, and electric fields are expressed in units of $h_1$, $\hbar/h_1$, $eh_1/(\hbar a)$, and $h_1/(ea)$, respectively.
For $h_1 = 1\ \mathrm{eV}$ and $a = 1\ \mathrm{nm}$, these read $\hbar/h_1 = 0.66\ \mathrm{fs}$, $eh_1/(\hbar a) = 2.4\ \mathrm{kA\, cm^{-1}}$, and $h_1/(ea) = 10\ \mathrm{MV\, cm^{-1}}$.

\section{Results} \label{sec:results}
In this section, we give an overview of the equilibrium properties of the four-sublattice scalar chiral state.
Then, we show the numerical results of the HHG when linearly polarized light is applied, and discuss how geometrical effects manifest themselves.
Hereafter, we focus mainly on cases where the Kondo coupling is $J_{\mathrm{K}} = 3$, and the electron number density is $n_{\mathrm{e}} = 0.5$ (half filling) \footnote{In a minimal ferromagnetic Kondo lattice model defined in Eq.~\eqref{eq:fklm} only with the nearest neighbor transfer integral, the scalar chiral state is not stable at $n_{\mathrm{e}} = 0.5$ \cite{Akagi2010}.
Nevertheless, we can stabilize the scalar chiral state by introducing superexchange interactions \cite{Akagi2010, Akagi2011, Akagi2015} or Dzyaloshinskii--Moriya interactions between localized spins, although such magnetic interactions do not directly affect the electronic structure.
Since we are currently interested in HHG in the presence of topological spin textures, we do not delve into the mechanisms stabilizing the scalar chiral state \cite{Akagi2012}, and instead, we fix the localized spins to analyze the real-time evolution of electrons.}.
At half filling, the optical gap increases proportionally with $J_{\mathrm{K}}$, which enables numerical analysis at optical frequencies that are sufficiently small relative to the gap, suppressing the excited electron density in the conduction bands.
Additionally, despite the absence of the dc Hall effect at $n_{\mathrm{e}} = 0.5$, optical transverse responses due to the nonzero Berry curvature can be observed, as shown in the following sections.

\subsection{Equilibrium properties} \label{sec:equil}

\begin{figure*}[t]\centering
\includegraphics[scale=1]{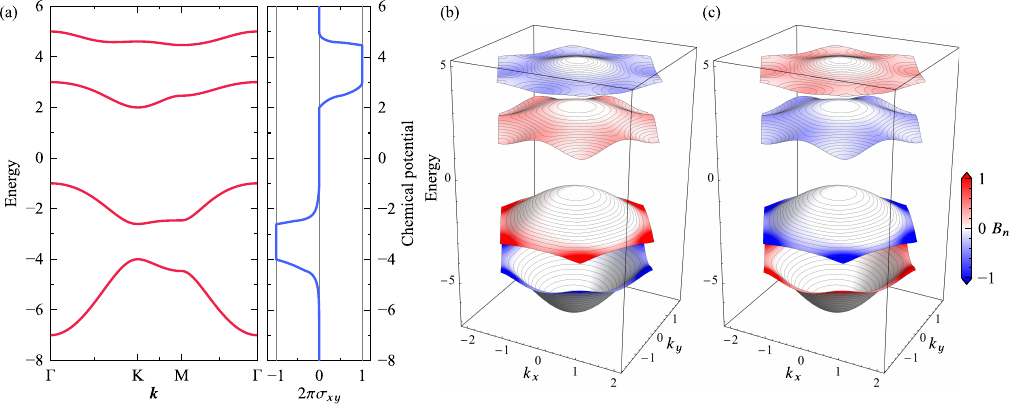}
\caption{(a)~Energy band structure $\varepsilon_n(\bm{k})$ (left), and the chemical potential dependence of dc Hall conductivity $2\pi \sigma_{xy}(0)$ at zero temperature (right), in the four-sublattice scalar chiral state.
The conductivity is calculated using Eqs.~\eqref{eq:conductivity}--\eqref{eq:kernel} with $\eta = 0.01$.
[(b) and (c)]~Berry curvature $B_n(\bm{k})$ plotted on the energy-band surfaces, for (b) $\chi = +\chi_{0}$ and (c) $\chi = -\chi_{0}$.}
\label{fig:bands_j3}
\end{figure*}

Figure~\ref{fig:bands_j3}(a) displays the energy band structure $\varepsilon_n(\bm{k})$ in the magnetic BZ alongside the dc Hall conductivity as a function of the chemical potential.
We observe four doubly degenerated bands in the magnetic BZ.
The dc Hall conductivity $\sigma_{xy}(\omega=0)$ exhibits a quantized value of $\pm e^2/(2\pi \hbar)$ when the electrons are at quarter or three quarter filling, as initially pointed out in Refs.\ \cite{Martin2008, Akagi2010}.
Since time reversal symmetry is broken while spatial inversion symmetry is preserved, the Berry curvature satisfying $\bm{B}_n(-\bm{k}) = \bm{B}_n(\bm{k})$ can appear in the scalar chiral state.
In Figs.~\ref{fig:bands_j3}(b) and \ref{fig:bands_j3}(c), we present the Berry curvature $B_n(\bm{k})$ on the energy-band surfaces for $\chi = +\chi_{0}$ and $\chi = -\chi_{0}$.
For the lower two doubly degenerated bands, the Berry curvature takes on large values at and in the vicinity of the K point, while for the upper two bands, the Berry curvature appears more dispersed throughout the BZ.
Notably, the sign of the Berry curvature is reversed by changing the sign of $\chi$ without affecting the energy bands $\varepsilon_n(\bm{k})$.
The Chern number defined in Eq.~\eqref{eq:chern} is numerically confirmed to be $\mathcal{C}_1 = -1$, $\mathcal{C}_2 = +1$, $\mathcal{C}_3 = +1$, and $\mathcal{C}_4 = -1$ from the bottom to the top band when $\chi = +\chi_{0}$, and their signs are reversed for $\chi = -\chi_{0}$, as is consistent with the chemical potential dependence of the dc Hall conductivity.
It should be emphasized that since the sign of scalar chirality alters only the sign of the Berry curvature, if the high harmonic spectrum differs depending on the chirality's sign, such a difference should be attributed not to the energy bands $\varepsilon_n(\bm{k})$ but to a purely geometrical effect originating from the Berry curvatures $\bm{B}_n(\bm{k})$.

\begin{figure}[t]\centering
\includegraphics[scale=1]{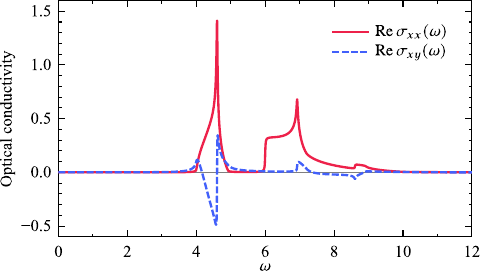}
\caption{Real part of optical conductivity in the scalar chiral state with $\chi = +\chi_{0}$ at zero temperature.
The parameters are set to $J_{\mathrm{K}}=3$, $n_{\mathrm{e}} = 0.5$, and $\eta = 0.01$.}
\label{fig:conductivity_j3}
\end{figure}

In the case of $J_{\mathrm{K}} = 3$ and $n_{\mathrm{e}} = 0.5$, the longitudinal component of the optical conductivity, $\sigma_{xx}(\omega)$, and the transverse component, $\sigma_{xy}(\omega)$, are shown in Fig.~\ref{fig:conductivity_j3}.
Although the direct optical band gap of $\varepsilon_3(\bm{k}) - \varepsilon_2(\bm{k}) = 4$ is at the $\Gamma$ point, there, the transition dipole moment proportional to $\bm{v}(\bm{k})$ is zero; a significant absorption peak can be seen in $\sigma_{xx}(\omega)$ at $\omega = 4.6$, corresponding to the interband transition at the K point.
At half filling, since the sum of the Chern numbers of the occupied bands is zero, the transverse conductivity $\sigma_{xy}(\omega)$ vanishes at $\omega = 0$, indicating no dc Hall effect.
Nonetheless, for $\omega \gtrsim 4$, nonzero $\sigma_{xy}(\omega)$ arises owing to interband transitions, and the sign of $\sigma_{xy}(\omega)$ also depends on the sign of scalar chirality.
This can be observed through linear magneto-optical effects, as discussed in Ref.\ \cite{Feng2020k}.
Even beyond such a linear and perturbative regime, given that the electromagnetic radiation intensity is determined by the expectation value of a one-body electric current operator, we anticipate transverse HHG dependent on scalar chirality.

\subsection{Real-time dynamics and HHG} \label{sec:hhg}

In this section, we examine the real-time dynamics when a continuous wave [Eq.~\eqref{eq:cw}] is applied, and discuss the characteristics of the resulting high harmonic spectrum.
First, we consider the case where linearly polarized light parallel to the $x$ axis (i.e., $\psi = 0$) is irradiated.
The relaxation rate and the optical frequency are set to $\varGamma = 0.1$ and $\varOmega = 10\times 2\pi/500 = 0.126$, respectively, with the latter being sufficiently smaller than the optical gap.
The ramp time in Eq.~\eqref{eq:cw} and the phase in Eq.~\eqref{eq:lp} are chosen as $\tau = 6$ and $\phi = 0$, respectively, which do not affect the high harmonic spectrum in a steady state.

\begin{figure}[t]\centering
\includegraphics[scale=1]{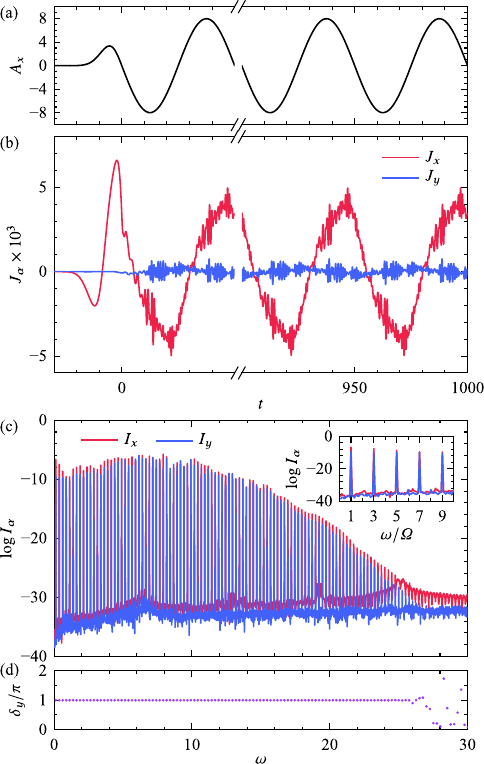}
\caption{[(a) and (b)]~Temporal profiles of (a) the vector potential, $A_x(t)$, and (b) the electric currents, $J_{x}(t)$ and $J_{y}(t)$.
(c)~Power spectra of the electric currents, $I_{x}(\omega)$ and $I_{y}(\omega)$.
The inset shows the intensity of low-order harmonics.
(d)~Phase difference in the transverse current $J_y(\omega)$ between states with opposite chiralities, $\delta_y = \arg J_y(\omega)\vert_{\chi = -\chi_{0}} - \arg J_y(\omega)\vert_{\chi = +\chi_{0}}$, for odd-order harmonics.
The electric field amplitude, frequency, polarization angle, and ramp time are set to $F_0 = 1$, $\varOmega = 0.126$, $\psi = 0$, and $\tau = 6$, respectively.}
\label{fig:timeprofile_spectrum_j3}
\end{figure}

We show the temporal profiles of the applied vector potential $\bm{A}(t) = (A_x(t), 0)$ and the electric current density $\bm{J}(t) = (J_x(t), J_y(t))$ in Figs.~\ref{fig:timeprofile_spectrum_j3}(a) and \ref{fig:timeprofile_spectrum_j3}(b), respectively, when the electric field amplitude is $F_0 = 1$.
Given the optical period of $T = 2\pi/\varOmega = 50$, the system quickly reaches a steady state after a few optical cycles (on a time scale of the order of $\varGamma^{-1} = 10$), where not only the longitudinal current $J_x$ but also the transverse current $J_y$ is induced by the vector potential parallel to the $x$ axis.
Although $J_y$ is less intense than $J_x$, its high-frequency oscillatory components are of similar magnitude to those of $J_x$.
As becomes clear from the subsequent discussion related to Fig.~\ref{fig:timeprofile_spectrum_j3}(d), this transverse response is due to $\sigma_{xy}\ (\neq 0)$ arising from scalar chirality, and as shown in Appendix~\ref{sec:120neel}, it does not occur in the $120^{\mathrm{\circ}}$ N\'eel state when linear polarization is along a high symmetric direction such as $\psi = 0$ and $\pi/6$.

High harmonic spectrum $I(\omega)$ can be obtained using the Fourier transformation of $\bm{J}(t)$.
In this study, we extracted real-time data for $500 < t \leq 1000$, considering the system to have reached the steady state before $t = 500$, and applied the fast Fourier transformation (FFT) to $(1000-500)/\delta t$ data points.
The optical frequency of $\varOmega = 10\times 2\pi/500$ is consistent with this number of data points, so that the FFT results include data points at frequencies that are exact multiples of $\varOmega$.

Figure~\ref{fig:timeprofile_spectrum_j3}(c) displays the intensities of the longitudinal and the transverse response, $I_x(\omega)$ and $I_y(\omega)$, respectively, obtained from $J_x(t)$ and $J_y(t)$ shown in Fig.~\ref{fig:timeprofile_spectrum_j3}(b).
Since the optical frequency is $\varOmega = 0.126$, which is less than $1/30$ of the optical gap, the high harmonic spectrum up to approximately the $200$th order is observed to be clearly separated from a background of $\lesssim 10^{-32}$.
Furthermore, as the optical period and the number of data points used for FFT are consistent, sharp peaks appear only at frequencies that are integer multiples of $\varOmega$ as shown in the inset of Fig.~\ref{fig:timeprofile_spectrum_j3}(c).
These high harmonic peaks are observed at odd orders, while the even-order harmonics disappear because of the presence of spatial inversion symmetry.
Overall, although the intensity of the transverse component, $I_y$, is several orders of magnitude lower than that of the longitudinal component, $I_x$, they appear in the same frequency range.
Up to about $\omega \lesssim 10$, a plateau appears in the spectrum, which roughly agrees with the frequency range where the optical conductivity is nonzero (see Fig.~\ref{fig:conductivity_j3}); this is a characteristic widely observed in the HHG in the nonperturbative regime.

\begin{figure*}[t]\centering
\includegraphics[scale=1]{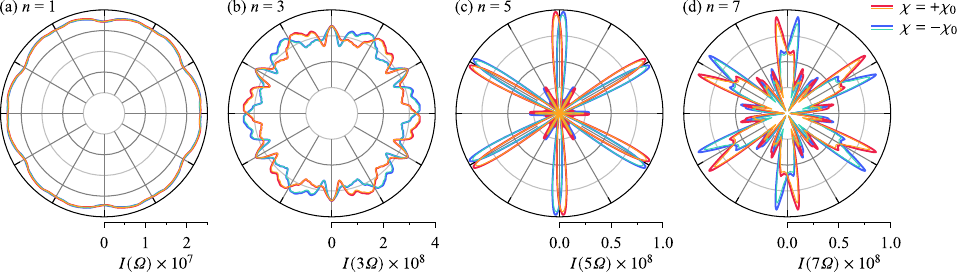}
\caption{[(a)--(d)]~Polarization angle dependence of the $n$th harmonic intensity $I(n\varOmega)$ for $\chi = +\chi_{0}$ (red) and $\chi = -\chi_{0}$ (blue).
The thin curves show the longitudinal component $I_\parallel(n\varOmega)$ for $\chi = +\chi_{0}$ (orange) and $\chi = -\chi_{0}$ (cyan).
The electric field amplitude and frequency are set to $F_0 = 1$ and $\varOmega = 0.126$, respectively.}
\label{fig:lpdirection_j3}
\end{figure*}

Here, we discuss how the transverse response $J_y$ changes with respect to the sign of scalar chirality.
We confirmed that, for $\chi = -\chi_{0}$, the power spectrum $I_y(\omega)$ is exactly the same as in the case of $\chi = +\chi_{0}$ shown in Fig.~\ref{fig:timeprofile_spectrum_j3}(c)~\footnote{As shown in Fig.~\ref{fig:lpdirection_j3}, the power spectrum also differs with the chirality sign for other polarization angles.}.
However, a difference is observed in the phase spectrum.
Figure~\ref{fig:timeprofile_spectrum_j3}(d) shows the difference in the phase component of $J_y(\omega)$, defined by $\delta_y = \arg J_y(\omega)\vert_{\chi = -\chi_{0}} - \arg J_y(\omega)\vert_{\chi = +\chi_{0}}$, for odd-order harmonics between the cases of $\chi = +\chi_{0}$ and $\chi = -\chi_{0}$.
As clearly seen in Fig.~\ref{fig:timeprofile_spectrum_j3}(d), the transverse component of the odd-order harmonics differs in phases by $\pi$ from each other.
This indicates that the sign of the transverse response in the scalar chiral state is inverted by time reversal, implying its association with the presence of scalar chirality, or the Berry curvature.

\begin{figure}[t]\centering
\includegraphics[scale=1]{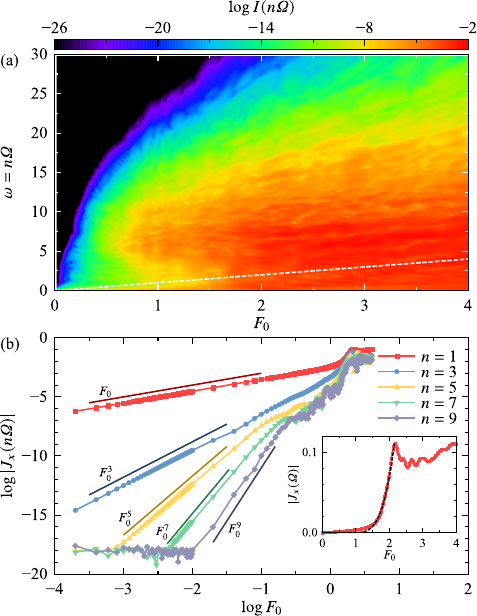}
\caption{(a)~Amplitude dependence of the power spectrum of odd-order harmonics, $I(n\varOmega)\ (n=1,3,\dots)$.
The dashed line indicates the upper bound of the Bloch oscillation frequency (see main text).
(b)~Fourier amplitude of low-order harmonics, $J_x(n\varOmega)$, as a function of $F_0$.
The inset shows $J_x(\varOmega)$ on a linear scale, with the dashed curve representing a fitted function, $|J_x(\varOmega)| = 52.02 \exp(-13.22/F_0)$.
The optical frequency and polarization angle are respectively set to $\varOmega = 0.126$ and $\psi = 0$ in (a) and (b).}
\label{fig:amplitude_j3}
\end{figure}

We show in Fig.~\ref{fig:amplitude_j3} the amplitude dependence of the high harmonic spectrum for $\chi = +\chi_0$.
Figure~\ref{fig:amplitude_j3}(a) is a color map displaying the intensity of odd-order harmonics, $I(n\varOmega)$, as a function of the electric field amplitude $F_0$.
Corresponding to the plateau region observed in Fig.~\ref{fig:timeprofile_spectrum_j3}(c), the intensity in the region of $\omega \lesssim 10$ is enhanced for $F_0 \gtrsim 1$.
The white dashed line in the figure indicates the upper bound of the Bloch oscillation frequency, which is $\omega = F_0$ in the case of $\psi = 0$.
The frequency domain mainly below this line can include contributions from intraband currents.

The detailed amplitude dependence of the low-order harmonics is plotted in Fig.~\ref{fig:amplitude_j3}(b).
For $F_0 \lesssim 0.1$, the Fourier amplitude of the $n$th harmonic, $J_x(n\varOmega)$, is proportional to the $n$th power of $F_0$, indicating that HHG is in the perturbative regime.
As $F_0$ increases, the higher-order harmonics begin to deviate from the perturbative regime, transitioning to the nonperturbative regime around $F_0 \sim 1$.
The inset of the figure plots the fundamental harmonic amplitude $\vert J_x(\varOmega) \vert$ on a linear scale.
This is well fitted by the exponential function $\exp(-F_{\mathrm{th}}/F_0)$ with $F_{\mathrm{th}} = 13.22$ indicated by the black dashed line, and the excited electron density exhibits similar behavior (not shown), suggesting that interband tunneling excitation dominates for $F_0 \gtrsim 1.5$.
Therefore, for $F_0 \lesssim 1.5$, tunneling excitation hardly occurs, and the geometrical effects on the tunneling probability discussed in Refs.\ \cite{Kitamura2020c, Takayoshi2021} can be considered negligible.

We examine the polarization angle $\psi$ dependence of the $n$th harmonic intensity for $F_0 = 1$, as shown in Fig.~\ref{fig:lpdirection_j3}.
The red and blue lines correspond to the cases of $\chi = +\chi_{0}$ and $\chi = -\chi_{0}$, respectively.
For the first-order harmonic, the difference due to the chirality sign is almost negligible, and it is approximately independent of the polarization angle.
This partially inherits the property that, in the current system with sixfold symmetry, a linear optical response exhibits continuous rotational symmetry (see Appendix~\ref{sec:perturbative} for details).
On the other hand, for the third and higher harmonics, not only does a significant dependence on the incident polarization angle appear, but clear differences are observed depending on the sign of scalar chirality.
As previously mentioned, the chirality sign only changes the sign of the Berry curvature and does not alter the energy band structure; hence, this chirality dependence is attributed to purely geometrical effects.

The polarization angle dependence reflecting the sign of scalar chirality is naively expected to arise from the anomalous velocity of intraband currents, as discussed in the literature \cite{Liu2016c, Schmid2021, Lv2021, Luu2018b, Bai2020a, Silva2019a, Lou2021} for systems where spatial inversion symmetry is broken.
The intraband current carried by an electron with momentum $\bm{k}$ in the $n$th band is proportional to
\begin{align}
\dot{\bm{r}}_n = \partial_{\bm{k}} \varepsilon_n(\bm{k}) + \bm{F}(t) \times \bm{B}_n(\bm{k}),
\label{eq:r_dot}
\end{align}
where $\bm{r}$ is the position of the electron, and the second term is called the anomalous velocity.
Note that the anomalous velocity term always produces a current perpendicular to the electric field $\bm{F}$.
Therefore, to extract the contribution of the transverse response, the power spectrum $I(\omega)$ is decomposed into components parallel and perpendicular to $\bm{F}$, denoted by $I_\parallel(\omega)$ and $I_\perp(\omega)$, respectively.
These are related to $I_x(\omega)$ and $I_y(\omega)$ through the relations:
\begin{align}
I_\parallel &= I_x \cos^2\psi + 2\sqrt{I_x I_y} \cos\psi \sin\psi \cos\delta + I_y \sin^2\psi, \\
I_\perp &= I_x \sin^2\psi -2 \sqrt{I_x I_y} \cos\psi \sin\psi \cos\delta + I_y \cos^2\psi,
\end{align}
with $\delta(\omega) = \arg J_y(\omega) - \arg J_x(\omega)$.
The thin curves in Fig.~\ref{fig:lpdirection_j3} show the polarization angle dependence of the intensity of the parallel component $I_\parallel(\omega) = I(\omega) - I_\perp(\omega)$.
Contrary to expectation, for any $\psi$, we observe that $I_\parallel \approx I$, indicating that the anomalous velocity (and a component of $\partial_{\bm{k}} \varepsilon_n(\bm{k})$ that is perpendicular to $\bm{k}(t) = \bm{k}-\bm{A}(t)$) in the intraband current cannot explain the observed dependence on the sign of scalar chirality.
Thus, in the following section, we consider interband currents associated with the recombination of electron--hole pairs.

\subsection{Electron--hole dynamics in real space} \label{sec:trajectory}
In the previous section, as shown in Fig.~\ref{fig:lpdirection_j3}, it was revealed that the polarization angle dependence of harmonic intensity changes with the sign of scalar chirality, and that it is mostly due to the contribution of an electric current component parallel to the electric field.
Since it is currently difficult to fully understand this cause microscopically, in this section, we discuss interband currents by analyzing a real-space trajectory of an electron--hole pair excited at a wave-number point $\bm{k} = \bm{k}_0$.

At half filling, optical driving primarily excites electrons into the third lowest band $\varepsilon_3(\bm{k})$, while creating holes in the second band $\varepsilon_2(\bm{k})$.
Interband currents are induced when these electrons and holes recombine.
In the saddle-point approximation \cite{Corkum1993, Lewenstein1994, Vampa2015a, Nayak2019a, Imai2022}, this condition is expressed as $\Vert \delta\bm{r} \Vert = 0$, where $\delta\bm{r}$ represents the relative displacement of the electron--hole pair excited at time $t = 0$.
This displacement is given by
\begin{align}
\delta\bm{r}(t) = \int_0^t \delta\dot{\bm{r}}(t')\, \mathrm{d}t',
\end{align}
where $\delta\dot{\bm{r}}$ denotes the relative velocity of the electron--hole pair with momentum $\bm{k}$:
\begin{align}
\delta\dot{\bm{r}}
&= \partial_{\bm{k}}[\varepsilon_3(\bm{k}) - \varepsilon_2(\bm{k})]
+ \bm{F}(t) \times [\bm{B}_3(\bm{k}) - \bm{B}_2(\bm{k})].
\label{eq:dr_dot}
\end{align}
The optical vector potential is introduced through the Peierls substitution: $\bm{k}(t) = \bm{k}_0 + \bm{A}(0) - \bm{A}(t)$ with $\bm{A}(t) = -(\bm{F}_0/\varOmega) \sin(\varOmega t - \phi)$.
Therefore, by finding the phase $\phi$ for which $\Vert \delta\bm{r} \Vert = 0$ at time $t > 0$, we can determine the real-space trajectory of the electron--hole pair until recombination.

\begin{figure*}[t]\centering
\includegraphics[scale=1]{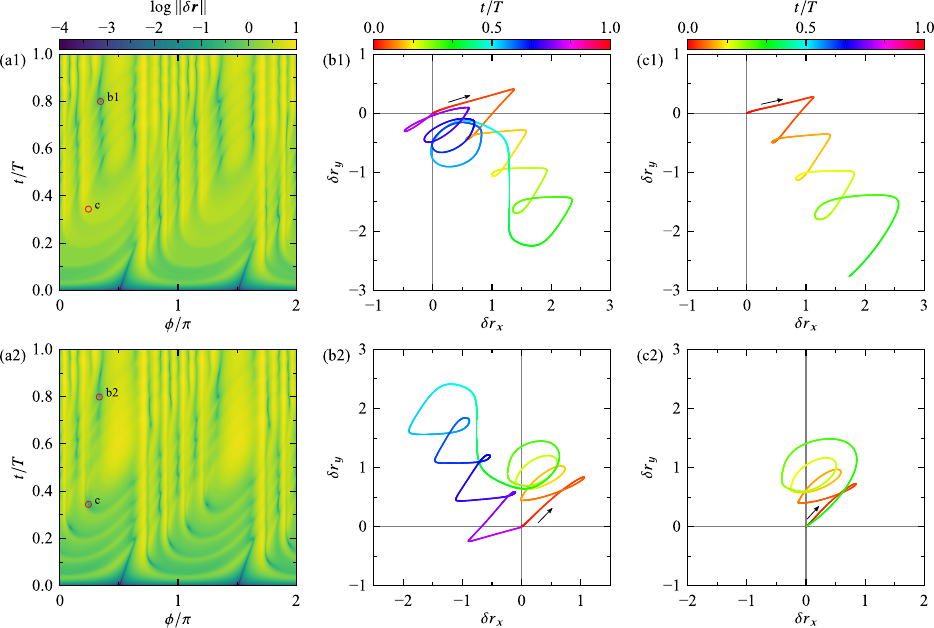}
\caption{Real-space dynamics of electron--hole pair excited at $t = 0$ with $\bm{k} = (0,0)$, for $\chi = +\chi_{0}$ (top) and $\chi = -\chi_{0}$ (bottom).
The electric field amplitude, frequency, polarization angle are set to $F_0 = 1$, $\varOmega = 0.126$, and $\psi = 32^{\mathrm{\circ}}$, respectively.
(a)~Color map of the norm of the relative displacement, $\Vert \delta \bm{r} \Vert$, in the $\phi$-$t$ plane.
(b)~Real-space trajectory of $\delta\bm{r}$ until recombination for an initial phase indicated by the red circle, ``b1'' and ``b2'', in (a).
(c)~Same as (b) but for a different initial phase indicated by ``c'' in (a); point ``c'' in (a1) is not a local minimum of $\Vert \delta \bm{r} \Vert$, and thus in (c1), the pair does not recombine (see main text).}
\label{fig:trajectory}
\end{figure*}

Analyzing all trajectories of electron--hole pairs for every $\bm{k}_0$ would only complicate the problem.
Thus, here, we specifically show representative cases for a pair excited at $\bm{k}_0 = (0,0)$, that is, at the $\Gamma$ point.
Figure~\ref{fig:trajectory}(a) displays the relative displacement $\Vert \delta\bm{r} \Vert$ on a logarithmic scale as a function of the initial phase $\phi$ and time $t$, with Fig.~\ref{fig:trajectory}(a1) for $\chi = +\chi_{0}$ and Fig.~\ref{fig:trajectory}(a2) for $\chi = -\chi_{0}$.
The polarization angle is set to $\psi = 32^{\mathrm{\circ}}$, for which the fifth-order harmonic intensity for $\chi = +\chi_{0}$ nearly reaches its maximum [see Fig.~\ref{fig:lpdirection_j3}(c)].
Overall, both cases exhibit similar behaviors, but, reflecting the sign of scalar chirality (i.e., the sign of the Berry curvature), the details differ.
There are specifically two cases in which the electron--hole pair can recombine: (i) for both $\chi = +\chi_{0}$ and $\chi = -\chi_{0}$, the pair recombines at almost the same phase $\phi$ and time $t$ (indicated by points ``b1'' and ``b2''), and (ii) the pair recombines only for either $\chi = +\chi_{0}$ or $\chi = -\chi_{0}$ (indicated by point ``c'').
We discuss these two cases in detail.

\begin{figure}[t]\centering
\includegraphics[scale=1]{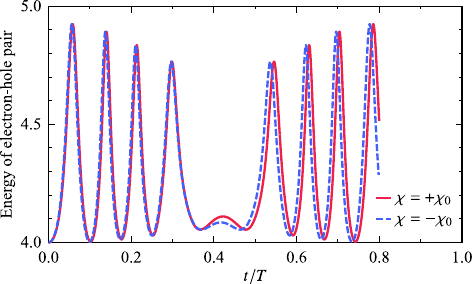}
\caption{Temporal profiles of the energy of electron--hole pair until the recombination, with the initial phases indicated by ``b1'' for $\chi = +\chi_{0}$ and by ``b2'' for $\chi = -\chi_{0}$ in Figs.~\ref{fig:trajectory}(a1) and \ref{fig:trajectory}(a2) (see main text).
The other parameters are the same as those in Fig.~\ref{fig:trajectory}.}
\label{fig:trajectory_energy}
\end{figure}

\textit{Case (i).}
The phase and the recombination time of points ``b1'' and ``b2'' in Figs.~\ref{fig:trajectory}(a1) and \ref{fig:trajectory}(a2) are $\phi/\pi = 0.344$ and $t/T= 0.800$ for $\chi = +\chi_{0}$, and $\phi/\pi = 0.333$ and $t/T= 0.799$ for $\chi = -\chi_{0}$.
Here, $T = 2\pi/\varOmega = 50$ represents the optical period.
The corresponding trajectories in real space are shown in Figs.~\ref{fig:trajectory}(b1) and \ref{fig:trajectory}(b2).
Although these trajectories are close to what would be expected if the time was reversed for the other, the contribution from the anomalous velocity term modifies the conditions for recombination.
This results in a slight difference in a recombination energy.
Figure~\ref{fig:trajectory_energy} shows the temporal profile of the electron--hole pair's energy, $\varepsilon_3\text{\textbf{(}}\bm{k}(t)\text{\textbf{)}} - \varepsilon_2\text{\textbf{(}}\bm{k}(t)\text{\textbf{)}}$, where we in fact observe the slight difference.
Thus, in this case, while the electron--hole pair recombines for both $\chi = +\chi_{0}$ and $\chi = -\chi_{0}$, their different recombination energies at $t\approx 0.8$ yield a different harmonic intensity.

\textit{Case (ii).}
When $\phi/\pi = 0.2416$ and $t/T = 0.344$, indicated as ``c'' in Fig.~\ref{fig:trajectory}(a), pair recombination occurs only for $\chi = -\chi_{0}$.
The trajectories for this case are shown in Figs.~\ref{fig:trajectory}(c1) and \ref{fig:trajectory}(c2).
For $\chi = +\chi_{0}$, the electron--hole pair does not return to the coordinate origin, and thus, this pair does not contribute to interband currents.

From the two cases above, the reason why the dependence of the chirality sign, as shown in Fig.~\ref{fig:lpdirection_j3}, appears as a longitudinal response can be inferred to be due to the difference in the dynamics of electron--hole pairs in real space.
This difference is caused by the anomalous velocity, which also changes the recombination conditions.
Furthermore, even in a case where recombination occur for both $\chi = +\chi_{0}$ and $\chi = -\chi_{0}$, the difference in the recombination energy results in variations in harmonic intensity.
However, it is also important to note that the analysis conducted here is significantly simplified and does not consider crucial factors such as temporal changes in the carrier density and interference with other pairs excited at different $\bm{k}$'s, necessitating more comprehensive analyses as conducted in Ref.\ \cite{Uzan-Narovlansky2024} in future work.

\subsection{HHG with parameters for real materials} \label{sec:material}

In the previous sections, we discussed the case where $J_{\mathrm{K}} = 3$ and $n_{\mathrm{e}} = 0.5$.
Recently, some experiments reported that the four-sublattice scalar chiral state is realized in $\mathrm{CoTa_3S_6}$ and $\mathrm{CoNb_3S_6}$ \cite{Takagi2023, Park2023, Park2024}.
Here, we discuss the high harmonic spectrum and its polarization angle dependence for parameters close to these materials, with $J_{\mathrm{K}} = 0.4$ and $n_{\mathrm{e}} = 0.75$.
We will see that the aforementioned conclusion regarding the dominance of the longitudinal response depending on the chirality sign also holds in this case.

\begin{figure}[t]\centering
\includegraphics[scale=1]{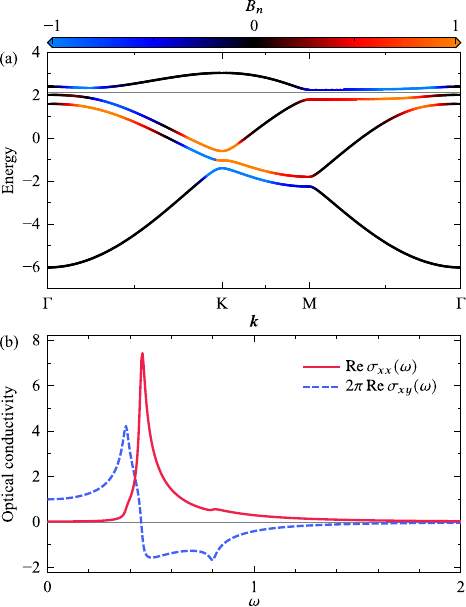}
\caption{(a)~Energy band structure $\varepsilon_n(\bm{k})$ for $J_{\mathrm{K}} = 0.4\ (\chi = +\chi_{0})$.
The color variation along the lines indicates the Berry curvature $B_n(\bm{k})$.
The gray horizontal line shows the chemical potential $\mu = 2.137$ for $n_{\mathrm{e}} = 0.75$.
(b)~Real part of optical conductivity in the scalar chiral state with $\chi = +\chi_{0}$ at zero temperature.
The parameters are set to $J_{\mathrm{K}}=0.4$, $n_{\mathrm{e}} = 0.75$, and $\eta = 0.01$.}
\label{fig:bands_j04}
\end{figure}

Before moving on to the discussion of HHG, we present the equilibrium properties.
Figure~\ref{fig:bands_j04}(a) shows the energy band structure and Berry curvature in the ground state for $J_{\mathrm{K}} = 0.4$.
Only at three-quarter filling ($n_{\mathrm{e}} = 0.75$), the ground state is insulating.
The optical conductivity is shown in Fig.~\ref{fig:bands_j04}(b).
A significant absorption peak in $\sigma_{xx}(\omega)$ is observed near $\omega = 0.42$, corresponding to the transition between the upper two bands on the $\Gamma$--M line.
Hereafter, the optical frequency will continue to be set at $\varOmega = 0.126$, which is still lower than the optical gap.
Furthermore, reflecting the Berry curvature, the optical Hall conductivity $\sigma_{xy}(\omega)$ also appears, and it reaches the quantized value of $+e^2/(2\pi \hbar)$ in the dc limit ($\omega \to 0$).

\begin{figure}[t]\centering
\includegraphics[scale=1]{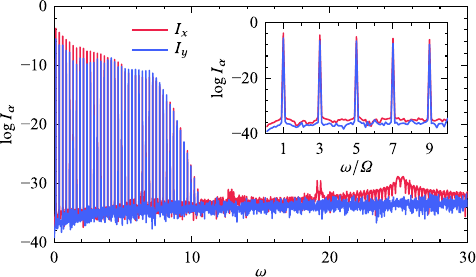}
\caption{Power spectra of the electric currents, $I_{x}(\omega)$ and $I_{y}(\omega)$.
The inset shows the intensity of low-order harmonics.
The parameters are set to $J_{\mathrm{K}}=0.4$, $n_{\mathrm{e}} = 0.75$, $F_0 = 0.1$, $\varOmega = 0.126$, and $N = 200^2$.}
\label{fig:spectrum_j04}
\end{figure}

In Fig.~\ref{fig:spectrum_j04}, we show the power spectrum $I_\alpha(\omega)$ for $F_0 = 0.1$ and $\varOmega = 0.126$.
Similarly to Fig.~\ref{fig:timeprofile_spectrum_j3}(c), the transverse response $I_y$ appears with the same order of magnitude as or several orders of magnitude smaller than the longitudinal response $I_x(\omega)$.
As the energy range from the bottom to the top band edge is approximately $9$ [see Fig.~\ref{fig:bands_j04}(a)], we observe the cutoff energy (i.e., the upper end of the plateau region) to be at the same energy $\sim 8$ in the spectrum.

\begin{figure}[t]\centering
\includegraphics[scale=1]{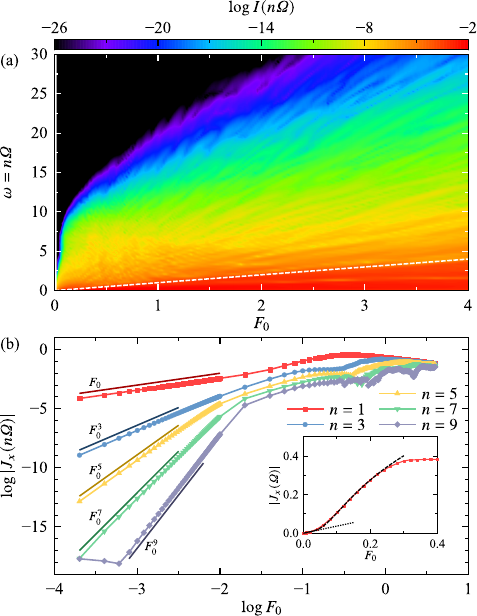}
\caption{(a)~Amplitude dependence of the power spectrum of odd-order harmonics, $I(n\varOmega)\ (n=1,3,\dots)$.
The dashed line indicates the upper bound of the Bloch oscillation frequency.
(b)~Fourier amplitude of low-order harmonics, $J_x(n\varOmega)$, as a function of $F_0$.
The inset shows $J_x(\varOmega)$ on a linear scale, with the dashed curve representing a fitted function, $|J_x(\varOmega)| = 0.3692 F_0 + 0.6032 \exp(-0.2177/F_0)$, and the dotted line being its linear component.
The parameters are set to $J_{\mathrm{K}}=0.4$, $n_{\mathrm{e}} = 0.75$, $\varOmega = 0.126$, $\psi = 0$, and $N = 100^2$ in (a) and (b).}
\label{fig:amplitude_j04}
\end{figure}

Figure~\ref{fig:amplitude_j04}(a) shows the color map of the intensity of odd-order harmonics for $\chi = +\chi_0$, as a function of the electric field amplitude.
Reflecting the observation in Fig.~\ref{fig:bands_j3}(b) that the optical gap is about an order of magnitude smaller than that in the case of $J_{\mathrm{K}} = 3$, the transition from the perturbative to the nonperturbative region occurs at a lower $F_0$.
The intensity $I(\omega)$ is particularly strong in the region of $\omega \lesssim 10$, consistent with the bandwidth of the electrons.
In addition, in the region below the white dashed line in Fig.~\ref{fig:amplitude_j04}(a), a significant contribution from intraband currents associated with the Bloch oscillation is also apparent.

The $F_0$ dependence of the harmonic intensities up to the ninth order is shown in Fig.~\ref{fig:amplitude_j04}(b).
For the fundamental harmonic ($n=1$), a deviation from the perturbative line $J_x(\varOmega) \propto F_0$ can be seen above $F_0 \sim 0.1$, and for higher harmonics, this deviation can be seen at a smaller $F_0$.
To consider the same situation as in the previous sections, the following discusses the polarization angle dependence for $F_0 = 0.1$.

\begin{figure*}[t]\centering
\includegraphics[scale=1]{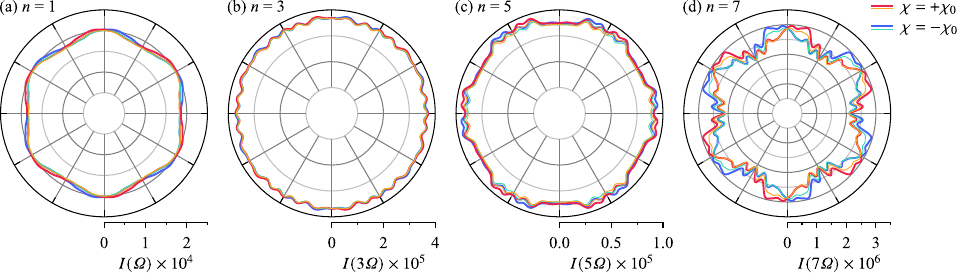}
\caption{[(a)--(d)]~Polarization angle dependence of the $n$th harmonic intensity $I(n\varOmega)$ for $\chi = +\chi_{0}$ (red) and $\chi = -\chi_{0}$ (blue).
The thin curves show the longitudinal component $I_\parallel(n\varOmega)$ for $\chi = +\chi_{0}$ (orange) and $\chi = -\chi_{0}$ (cyan).
The parameters are set to $J_{\mathrm{K}}=0.4$, $n_{\mathrm{e}} = 0.75$, $F_0 = 0.1$, and $\varOmega = 0.126$.}
\label{fig:lpdirection_j04}
\end{figure*}

We present the polarization angle dependence of harmonic intensity in Fig.~\ref{fig:lpdirection_j04}.
Similarly to the case of $J_{\mathrm{K}} = 3$ and $n_{\mathrm{e}} = 0.5$, the harmonic intensities depend on the polarization angle $\psi$, reflecting the sign of chirality.
However, for harmonics up to the fifth order at $F_0 = 0.1$, the significant angle dependence shown in Figs.~\ref{fig:lpdirection_j3}(c) and \ref{fig:lpdirection_j3}(d) are not observed.
Additionally, the thin dashed lines in the figure indicates the longitudinal intensity $I_\parallel$ parallel to the electric field $\bm{F}_0$, which, as in the previous case, satisfies $I_\parallel \approx I$, indicating the dominance of the longitudinal response depending on the chirality sign.
Therefore we consider that the results and discussions in the previous sections do not qualitatively depend on details such as model parameters or electron density.

\section{Discussion} \label{sec:discussion}
In Sec.~\ref{sec:results}, we have focused particularly on a case where the optical frequency is significantly lower than the energy gap.
Previous studies \cite{Liu2016c, Schmid2021, Lv2021, Luu2018b, Bai2020a, Silva2019a, Lou2021} have mainly discussed the effects of anomalous velocity in intraband currents; however, our results reveal that despite the dominance of the longitudinal response over the transverse response, the polarization angle dependence of harmonic intensity strongly reflects the sign of scalar chirality.
While this behavior might also be observed in systems with broken spatial inversion symmetry and nonzero Berry curvature, note that the sign of the Berry curvature can be easily switched by an external magnetic field in systems with broken time reversal symmetry.

It is also a natural question whether the anomalous velocity term in intraband currents (i.e., the transverse response) could dominate in the present system.
In Appendix~\ref{sec:near_resonant}, we show results on the polarization angle dependence in the case of near-resonant driving.
There, indeed, the transverse response can become comparable to or greater than the longitudinal response.
Besides, it is noteworthy that the longitudinal response still exhibits a dependence on the chirality sign.
In addition, Appendices~\ref{sec:circular} and \ref{sec:120neel} respectively present brief summaries of the high harmonic spectrum in the case with circular polarization driving, and of the polarization angle dependence of the harmonic intensities in the $120^{\mathrm{\circ}}$ N\'eel state, where the Berry curvature is zero.

As already mentioned, the linear optical Hall effect with topological spin textures has been discussed in the literature \cite{Hayashi2021, Sorn2021, Feng2020k, Kato2023, Li2024}.
In systems with sixfold symmetry like the one considered here, the linear conductivity exhibits continuous rotational symmetry, and thus shows no polarization angle dependence, unlike what is observed in Figs.~\ref{fig:lpdirection_j3} and \ref{fig:lpdirection_j04}.
Therefore, to verify the results presented in this paper, experiments need to be conducted on single crystals without grain boundaries.
Additionally, the scalar chiral state in $\mathrm{CoTa_3S_6}$ and $\mathrm{CoNb_3S_6}$ is metallic \cite{Takagi2023, Park2023, Park2024}, leading to the enhancement of the intraband-current response.
Thus, examining harmonics in a frequency range higher than the Bloch oscillation frequency would facilitate a clearer observation of the contribution from interband currents.

\section{Summary} \label{sec:summary}
In this paper, we numerically analyzed HHG arising from electrons in the spin scalar chiral state.
Reflecting the presence of the Berry curvature, the transverse response emerges, which is of the same order of magnitude as, or several orders of magnitude smaller than, the longitudinal response; its phase inversion depends on the sign of scalar chirality.
Furthermore, we observed a marked variation in harmonic intensity with respect to the incident polarization angle, dependent on the chirality sign, with the dominant component being the longitudinal response rather than the transverse one.
Since the anomalous velocity term in intraband currents produces only the transverse currents, this longitudinal response can be attributed to interband currents induced by the recombination of electron--hole pairs whose trajectories are modulated by the anomalous velocity.
This modulation changes the recombination energies of the pairs and thus can alter the spectrum of interband currents.
These results indicate that the magnetic structure with scalar chirality is, in fact, reflected in the high harmonic spectrum through the electron dynamics, which can be verified in experiments with materials such as $\mathrm{CoTa_3S_6}$ and $\mathrm{CoNb_3S_6}$, where the sign of scalar chirality can be switched by a magnetic field.
Further research is expected to extend to HHG and HSG in systems with other topological spin textures, such as skyrmion lattice and hedgehog lattice states.
Additionally, while the localized spins are fixed in this study, considering the dynamics resulting from the coupling between electrons and magnons would present an interesting direction \cite{Ono2023, Ono2017, Ono2018, Hattori2024}.

\begin{acknowledgments}
This work was supported by JSPS KAKENHI Grants No.\ JP23K13052, No.\ JP23K25805, No.\ JP24K00563, No.\ JP23K19030, No.\ JP23K25816, No.\ JP22K13998, and No.\ JP24K00546.
S.O. is supported by JST CREST Grants No.\ JPMJCR18T2 and No.\ JPMJCR19T3.
S.I. is supported by JST FOREST Grant No.\ JPMJFR2131.
Y.A. is supported by JST PRESTO Grant No.\ JPMJPR2251.
The numerical calculations were performed using the facilities of the Supercomputer Center, the Institute for Solid State Physics, the University of Tokyo.
\end{acknowledgments}

\appendix
\section{Near-resonant driving} \label{sec:near_resonant}
\begin{figure*}[t]\centering
\includegraphics[scale=1]{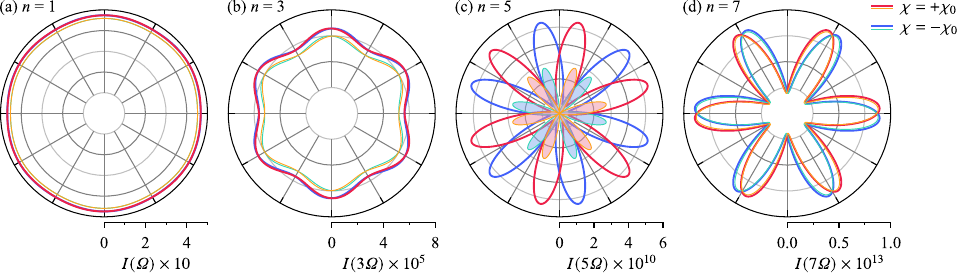}
\caption{[(a)--(d)]~Polarization angle dependence of the $n$th harmonic intensity $I(n\varOmega)$ for $\chi = +\chi_{0}$ (red) and $\chi = -\chi_{0}$ (blue).
The thin curves show the longitudinal component $I_\parallel(n\varOmega)$ for $\chi = +\chi_{0}$ (orange) and $\chi = -\chi_{0}$ (cyan), and $I_\parallel(5\varOmega)$ is additionally shaded for visibility.
The electric field amplitude and frequency are set to $F_0 = 1$ and $\varOmega = 4.02$, respectively.}
\label{fig:lpdirection_j3_resonant}
\end{figure*}

In the main text, we set the optical frequency to $\varOmega = 0.126$ and discussed the situation where it is significantly lower than the optical gap of $4$ for $J_{\mathrm{K}}=3$.
Here, we present in Fig.~\ref{fig:lpdirection_j3_resonant} the polarization angle dependence of harmonic intensity for a near-resonant case with $\varOmega = 320\times 2\pi/500 = 4.02$ and $F_0 = 1$.
Since $\varOmega$ is near resonant, the intensity of the fundamental harmonic is six orders of magnitude larger than that in Fig.~\ref{fig:lpdirection_j3}(a), but the angle dependence is small, suggesting that its deviation from the perturbative regime is small.
The higher order harmonics show a pronounced polarization angle dependence similar to the case of $\varOmega = 0.126$, and changes relative to the sign of scalar chirality can be similarly observed.
Among the higher order harmonics shown in Fig.~\ref{fig:lpdirection_j3_resonant}, the third- and seventh-order longitudinal response satisfies $I_\parallel \approx I$, but for the fifth harmonic, the transverse response $I_\perp$ becomes comparable to or greater than $I_\parallel$.
This large transverse response can be attributed to the anomalous velocity term in intraband currents, as discussed in the literature.
Nevertheless, $I_\parallel$ still clearly depends on the chirality sign, indicating that $I(\omega)$ contains interband-current contributions discussed in Sec.~\ref{sec:trajectory}.

\section{Circular polarization driving} \label{sec:circular}
Here, we briefly discuss HHG when circularly polarized light defined in Eq.~\eqref{eq:cp} is applied.
Before that, we present the relationship between the parameters of an ellipse and the electric current's amplitude and phase.
When the electric current corresponding to the $n$th harmonic is given by
\begin{align}
J_{\alpha}(t) = J_{0,\alpha} \cos(n\varOmega t - \phi_\alpha)
\end{align}
with $J_{0,\alpha} = \vert J_{\alpha}(n\varOmega) \vert$ and $\phi_\alpha = \arg J_{\alpha}(n\varOmega)$, the trajectory on the $J_x$-$J_y$ plane is an ellipse.
Its semi-major axis $J_+$ and semi-minor axis $J_-$ are respectively given by
\begin{align}
J_+ = \max\{\tilde{J}_x, \tilde{J}_y\}, \quad
J_- = \sgn(\delta) \min\{\tilde{J}_x, \tilde{J}_y\},
\end{align}
where $\delta = \phi_y-\phi_x$ represents the relative phase, $\sgn$ denotes a sign function, and $\tilde{J}_x$ and $\tilde{J}_y$ are defined by
\begin{align}
\tilde{J}_x
&= \vert \sin\delta\, \vert \left[ \frac{\cos^2\varphi}{J_{0,x}^2} - \frac{\sin2\varphi\cos\delta}{J_{0,x}J_{0,y}} + \frac{\sin^2\varphi}{J_{0,y}^2} \right]^{-\frac{1}{2}}, \\
\tilde{J}_y
&= \vert \sin\delta\, \vert \left[ \frac{\sin^2\varphi}{J_{0,x}^2} + \frac{\sin2\varphi\cos\delta}{J_{0,x}J_{0,y}} + \frac{\cos^2\varphi}{J_{0,y}^2} \right]^{-\frac{1}{2}}.
\end{align}
Here, $\varphi$ represents the inclination angle of the ellipse's major axis with respect to the $x$ axis, which is given by
\begin{align}
\varphi = \frac{1}{2} \arctan \frac{2J_{0,x} J_{0,y} \cos\delta}{J_{0,x}^2 - J_{0,y}^2}.
\end{align}
The ellipticity $\epsilon$ is defined by
\begin{align}
\epsilon = - \frac{J_{-}}{J_{+}},
\end{align}
such that it equals $+1$ for RCP and $-1$ for LCP.

\begin{figure}[b]\centering
\includegraphics[scale=1]{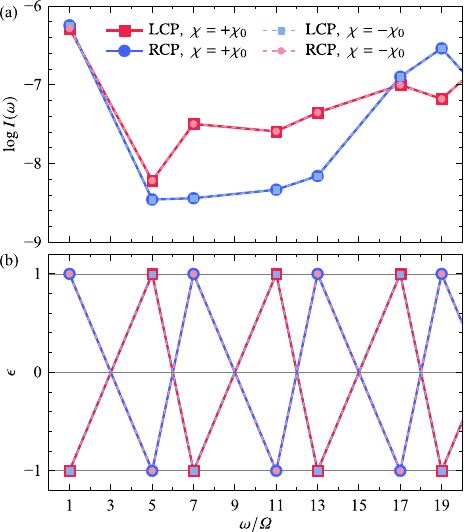}
\caption{(a)~Power spectrum $I(\omega)$ and (b)~ellipticity $\epsilon$, for $(6l\pm 1)$th-order harmonics.
RCP and LCP correspond to $\epsilon = +1$ and $\epsilon = -1$, respectively.
The parameters are set to $J_{\mathrm{K}}=3$, $n_{\mathrm{e}} = 0.5$, $F_0 = 0.1$, and $\varOmega = 0.126$.}
\label{fig:cp}
\end{figure}

Figure~\ref{fig:cp}(a) shows the power spectrum $I(\omega)$ for harmonics with intensity sufficiently separated from the background ($\lesssim 10^{-30}$).
It is established that the allowed harmonics for a given crystal symmetry and optical-field waveform are described by a theory of dynamical symmetry \cite{Alon1998, Liu2016, Neufeld2019, Nagai2020a, Ikeda2020a, Kanega2024}.
In the present system, which exhibits sixfold symmetry, only the $(6l\pm 1)$th-order harmonics ($l\in \mathbb{Z}$) are allowed when circularly polarized light is applied, and our results are in agreement with this theoretical prediction.
Additionally, Fig.~\ref{fig:cp}(b) demonstrates that the ellipticity $\epsilon$ of each harmonic is determined solely by the handedness of the circular polarization, regardless of the sign of chirality.
Furthermore, a kind of circular dichroism is observed; that is, for $\chi = +\chi_{0}$, the intensity of the fifth- to $13$th-order harmonics under LCP is more pronounced than those under RCP, and this difference in intensities is inverted when the sign of chirality is altered.

\section{Comparison with the 120${}^{\mathrm{\circ}}$ N\'eel state} \label{sec:120neel}
\begin{figure}[b]\centering
\includegraphics[scale=1]{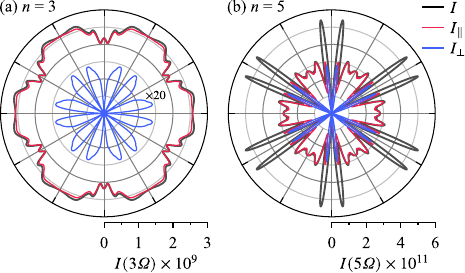}
\caption{Polarization angle dependence of the (a) third and (b) fifth harmonic intensity.
The red and blue lines represent the longitudinal ($I_\parallel$) and transverse ($I_\perp$) component, respectively.
For visibility, $I_\perp(3\varOmega)$ in (a) is multiplied by $20$.}
\label{fig:lpdirection_j3_3sub}
\end{figure}

The $120^{\mathrm{\circ}}$ N\'eel state exhibits a three-sublattice magnetic order, which is stabilized at $n_{\mathrm{e}} = 0.5$ in the present model \cite{Akagi2010}.
The vector of the $m$th sublattice spin is defined by $\bm{S}_m = ( \cos\theta_m, \sin \theta_m, 0)$ with $\theta_1 = 0$, $\theta_2 = 2\pi/3$, and $\theta_3 = -2\pi/3$.
Given this configuration, the electron system is invariant under the combination of a mirror reflection with respect to the $xy$ plane and the time reversal operation; thereby the Berry curvature $\bm{B}_n(\bm{k})$ satisfies $B_n^z(-k_x, -k_y, k_z) = -B_n^z(k_x, k_y, k_z)$ \cite{Suzuki2017b}, or in two dimensions, $B_n(-k_x, -k_y) = -B_n(k_x, k_y)$.
Additionally, the presence of spatial inversion symmetry imposes $B_n(-k_x, -k_y) = B_n(k_x, k_y)$.
Therefore the Berry curvature $B_n(\bm{k})$ turns out to be zero for any $\bm{k}$.
In the case with $J_{\mathrm{K}} = 3$ and $n_{\mathrm{e}} = 0.5$ as adopted in the main text, the ground state is insulating, and the optical gap is $2J_{\mathrm{K}} = 6$.

In Fig.~\ref{fig:lpdirection_j3_3sub}, we present the polarization angle dependence of the third- and fifth-order harmonic intensities for $F_0 = 1$ and $\varOmega = 0.126$, along with their longitudinal and transverse components.
Similarly to the case in the four-sublattice scalar chiral state, sixfold symmetric harmonic intensity is observed, with the fifth harmonic showing more pronounced angle dependence.
The transverse component $I_\perp$, shown in blue lines in the figure, appears except in the high symmetric directions such as $\psi = 0$ and $\psi = \pi/6$, and it does not depend on the sign of $J_{\mathrm{K}}$ unlike in the scalar chiral state.
Such transverse response can be attributed to the first term of Eq.~\eqref{eq:r_dot}, $\partial_{\bm{k}} \varepsilon_n(\bm{k})$, having components that are not parallel to the momentum $\bm{k}$.

\section{Nonlinear response in the perturbative regime} \label{sec:perturbative}
We derive the perturbative expressions for the harmonic intensity and discuss the polarization angle dependence.
In general, the $n$th-order response of the electric current $\bm{J}$ to the electric field $\bm{F}$ is given by
\begin{align}
J_\alpha^{(n)}(\omega)
&= \int_{-\infty}^{\infty} \frac{\mathrm{d}\omega_1 \cdots \mathrm{d}\omega_n}{(2\pi)^{n-1}} \delta(\omega_1 + \cdots + \omega_n - \omega) \notag \\
&\quad \times \sigma_{\alpha \alpha_1 \cdots \alpha_n}^{(n)}(\omega_1, \dots, \omega_n) F_{\alpha_1}(\omega_1) \cdots F_{\alpha_n}(\omega_n),
\label{eq:nth_response}
\end{align}
where $\sigma^{(n)}$ is the $n$th-order response function, $\alpha$ and $\alpha_{i=1,\dots, n}$ indicate Cartesian components (such as $x$ and $y$), and summation over repeated indices (Einstein summation convention) is assumed.
By definition, $\sigma^{(n)}$ is invariant under the permutation of $(\alpha_i, \omega_i) \leftrightarrow (\alpha_j, \omega_j)$, and thus, it is convenient to introduce the $n$th-order symmetrized response function:
\begin{align}
&\bar{\sigma}_{\alpha \alpha_1 \cdots \alpha_n}^{(n)}(\omega_1, \dots, \omega_n) \notag \\
&= \sum_{s \in \mathfrak{S}_n} \sigma_{\alpha \alpha_{s(1)} \cdots \alpha_{s(n)}}^{(n)}(\omega_{s(1)}, \dots, \omega_{s(n)}),
\label{eq:nth_response_sym}
\end{align}
where $\mathfrak{S}_n$ is the symmetric group of degree $n$.
For example, when $n=2$, Eq.~\eqref{eq:nth_response_sym} is written as $\bar{\sigma}_{\alpha \alpha_1 \alpha_2}^{(2)}(\omega_1, \omega_2) = \sigma_{\alpha \alpha_1 \alpha_2}^{(2)}(\omega_1, \omega_2) + \sigma_{\alpha \alpha_2 \alpha_1}^{(2)}(\omega_2, \omega_1)$.
Additionally, if the system is invariant under a symmetry operation represented by a unitary matrix $U$, the response function satisfies the relation:
\begin{align}
\sigma_{\alpha \alpha_1 \cdots \alpha_n}^{(n)} = U_{\alpha \beta} U_{\alpha_1 \beta_1} \cdots U_{\alpha_n \beta_n} \sigma_{\beta \beta_1 \cdots \beta_n}^{(n)},
\label{eq:nth_symmetry}
\end{align}
which reduces the number of independent nonzero components of $\sigma_{\alpha \alpha_1 \cdots \alpha_n}^{(n)}$.

Given that the present system preserves sixfold symmetry, we obtain from Eq.~\eqref{eq:nth_symmetry} the well-known relations,
\begin{align}
\sigma_{xx}^{(1)} = \sigma_{yy}^{(1)}, \quad
\sigma_{xy}^{(1)} = -\sigma_{yx}^{(1)}
\end{align}
for the first-order response.
Such relations can also be derived for higher-order responses, while we do not write them all out.
Instead, here we discuss the polarization angle dependence of the $n$th harmonic intensity, assuming the linearly polarized electric field,
\begin{align}
\bm{F}(\omega) = 2\pi \delta(\omega-\varOmega) \bm{F}_0, \quad
\bm{F}_0 = F_0 (\cos\psi, \sin\psi).
\label{eq:lp_electricfield}
\end{align}
Here, $F_0$ and $\varOmega$ represent the amplitude and frequency of the electric field, respectively, and $\psi$ denotes the polarization angle as in Eq.~\eqref{eq:lp}.
Under this assumption, the $n$th-order response in Eq.~\eqref{eq:nth_response} reduces to
\begin{align}
J_{\alpha}^{(n)}(\omega) &= 2\pi \delta(\omega-n\varOmega) \sigma_{\alpha \alpha_1 \cdots \alpha_n}^{(n)} F_{0,\alpha_1} \cdots F_{0,\alpha_n}, \\
J_{\alpha}^{(n)}(t) &= \int_{-\infty}^{\infty} \frac{\mathrm{d}\omega}{2\pi} \mathrm{e}^{-\mathrm{i}\omega t} J_{\alpha}^{(n)}(\omega) \notag \\
&= \mathrm{e}^{-\mathrm{i} n \varOmega t} \sigma_{\alpha \alpha_1 \cdots \alpha_n}^{(n)} F_{0,\alpha_1} \cdots F_{0,\alpha_n} \notag \\
&= \mathrm{e}^{-\mathrm{i} (n \varOmega t - \phi_{\alpha}^{(n)})} \bigl\vert J_{0,\alpha}^{(n)} \bigr\vert,
\end{align}
where
\begin{align}
J_{0,\alpha}^{(n)} = \sigma_{\alpha \alpha_1 \cdots \alpha_n}^{(n)} F_{0,\alpha_1} \cdots F_{0,\alpha_n}, \quad
\phi_{\alpha}^{(n)} = \arg J_{0,\alpha}^{(n)},
\label{eq:nth_amplitude}
\end{align}
and $\sigma_{\alpha \alpha_1 \cdots \alpha_n}^{(n)}$ is a shorthand for $\sigma_{\alpha \alpha_1 \cdots \alpha_n}^{(n)}(\varOmega, \dots, \varOmega)$.
Consequently, the $n$th-order harmonic intensity is given by
\begin{align}
I(n\varOmega) = (n\varOmega)^2 \left( \bigl\vert J_{0,x}^{(n)} \bigr\vert^2 + \bigl\vert J_{0,y}^{(n)} \bigr\vert^2 \right).
\label{eq:harmonic_intensity}
\end{align}

\begin{figure}[t]\centering
\includegraphics[scale=1]{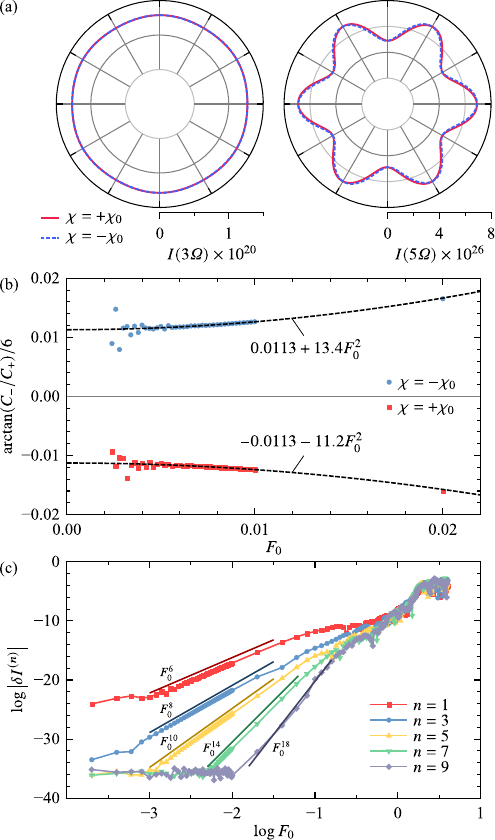}
\caption{(a)~Polarization angle dependence of the third (left) and fifth (right) harmonic intensity in the perturbative regime, $F_0 = 0.01$.
(b)~Amplitude dependence of the polarization angle at which the fifth harmonic intensity $I(5\varOmega)$ is maximized.
The dashed curves represent quadratic fits to the data points for $F_0 \in [0.006, 0.01]$.
(c)~Amplitude dependence of $\delta I^{(n)}$ in Eq.~\eqref{eq:diff_intensity} when $\chi = +\chi_0$.
The parameters are set to $J_{\mathrm{K}} = 3$, $n_{\mathrm{e}} = 0.5$, and $\varOmega = 0.126$ in (a)--(c).}
\label{fig:perturbative}
\end{figure}

By using Eqs.~\eqref{eq:nth_symmetry} and \eqref{eq:lp_electricfield}--\eqref{eq:harmonic_intensity}, we obtain an expression for the first-order harmonic intensity,
\begin{align}
I(\varOmega) \propto \left( \bigl\vert \sigma_{xx}^{(1)} \bigr\vert^2 + \bigl\vert \sigma_{xy}^{(1)} \bigr\vert^2 \right) F_0^2,
\label{eq:1st_intensity}
\end{align}
which is independent of $\psi$ as mentioned in the main text.
Similarly, the third harmonic intensity, written as
\begin{align}
I(3\varOmega) \propto \left( \bigl\vert \bar{\sigma}_{xxxx}^{(3)} \bigr\vert^2 + \bigl\vert \bar{\sigma}_{xxxy}^{(3)} \bigr\vert^2 \right) F_0^6,
\label{eq:3rd_intensity}
\end{align}
is also independent of $\psi$.
However, the fifth harmonic intensity turns out to be
\begin{align}
I(5\varOmega) &\propto \left[ C_0 + C_+ \cos(6\psi) + C_- \sin(6\psi) \right] F_0^{10},
\label{eq:5th_intensity}
\end{align}
which exhibits sixfold symmetry in the perturbative regime and reaches its maximum when $6\psi = \arctan(C_-/C_+) \bmod{2\pi}$.
Here, $C_0$, $C_+$, and $C_-$ are constants given by
\begin{align}
C_0 &= 117 \bigl\vert\bar{\sigma}_{yyyxxx}^{(5)}\bigr\vert^2 + 432 \bigl\vert\bar{\sigma}_{yyyxxx}^{(5)} \bar{\sigma}_{yyyyyx}^{(5)}\bigr\vert + 468 \bigl\vert\bar{\sigma}_{yyyyyx}^{(5)}\bigr\vert^2 \label{eq:5th_coefficient_c0} \notag \\
&\quad + 25\left( \bigl\vert\bar{\sigma}_{yyyyxx}^{(5)}\bigr\vert^2 + 8 \bigl\vert\bar{\sigma}_{yyyyxx}^{(5)} \bar{\sigma}_{yyyyyy}^{(5)}\bigr\vert + 52 \bigl\vert\bar{\sigma}_{yyyyyy}^{(5)}\bigr\vert^2 \right), \\
C_+ &= -45 \bigl\vert\bar{\sigma}_{yyyxxx}^{(5)}\bigr\vert^2 + 180 \bigl\vert\bar{\sigma}_{yyyyyx}^{(5)}\bigr\vert^2 \notag \\
&\quad + 25 \left( \bigl\vert\bar{\sigma}_{yyyyxx}^{(5)}\bigr\vert - 2\bigl\vert\bar{\sigma}_{yyyyyy}^{(5)}\bigr\vert \right) \left( \bigl\vert\bar{\sigma}_{yyyyxx}^{(5)}\bigr\vert + 10\bigl\vert\bar{\sigma}_{yyyyyy}^{(5)}\bigr\vert \right), \\
C_- &= -90 \bigl\vert\bar{\sigma}_{yyyxxx}^{(5)} \bar{\sigma}_{yyyyxx}^{(5)}\bigr\vert \notag \\
&\quad - 120 \bigl\vert\bar{\sigma}_{yyyyyx}^{(5)}\bigr\vert \left(\bigl\vert\bar{\sigma}_{yyyyxx}^{(5)}\bigr\vert - 5 \bigl\vert\bar{\sigma}_{yyyyyy}^{(5)}\bigr\vert \right). \label{eq:5th_coefficient_cm}
\end{align}
For the seventh harmonic intensity, we find that
\begin{align}
I(7\varOmega) &\propto [ D_0 + D_{1+} \cos(6\psi) + D_{1-} \sin(6\psi) \notag \\
&\quad + D_{2+} \cos(12\psi) + D_{2-} \sin(12\psi) ] F_0^{14},
\label{eq:7th_intensity}
\end{align}
where $D_0$, $D_{1\pm}$, and $D_{2\pm}$ are constants depending on $\sigma^{(7)}$.
Note that Eqs.~\eqref{eq:1st_intensity}--\eqref{eq:7th_intensity} are perturbative expressions valid in the limit of $F_0 \to 0$ and are derived solely from the sixfold symmetry, without making any additional assumptions regarding the electronic or magnetic structure.

Figure~\ref{fig:perturbative}(a) shows the polarization angle dependence of the third and fifth harmonics for $F_0 = 0.01$.
At this value of $F_0$, the response is within the perturbative regime, as evidenced by Fig.~\ref{fig:amplitude_j3}(b).
In fact, the numerical results exhibit continuous rotational symmetry for the third harmonic and sixfold symmetry for the fifth harmonic, which are consistent with Eqs.~\eqref{eq:3rd_intensity} and \eqref{eq:5th_intensity}.

We also notice that, in Fig.~\ref{fig:perturbative}(a), while the third harmonic intensity becomes independent of the chirality sign as $F_0 \to 0$, the fifth harmonic intensities show a slight difference between the cases of $\chi = +\chi_0$ and $-\chi_0$.
When the electric-field amplitude is finite, $F_0$-dependent terms enter Eqs.~\eqref{eq:5th_coefficient_c0}--\eqref{eq:5th_coefficient_cm} through higher-order processes.
Thus, to clarify whether the slight chirality-sign dependence observed in the fifth harmonic persists in the limit of $F_0 \to 0$, it is necessary to examine the dependence on $F_0$.
In Fig.~\ref{fig:perturbative}(b), the polarization angle at which $I(5\varOmega)$ is maximized is plotted as a function of $F_0$, and it deviates from the $x$-axis by $\mp 0.0113\ \mathrm{rad} = \mp 0.65^\circ$ for $\chi=\pm \chi_0$ in the limit of $F_0 \to 0$.
Since this deviation is quite small, the significant chirality-sign dependence observed in Figs.~\ref{fig:lpdirection_j3}, \ref{fig:lpdirection_j04}, and \ref{fig:lpdirection_j3_resonant}, as well as the pronounced polarization angle dependence with a nodelike structure, is likely enhanced by nonperturbative effects.

To further validate that the polarization angle dependence observed in Fig.~\ref{fig:perturbative}(a) can be described perturbatively, we consider the difference between the harmonic intensities for $\psi = 0$ and $\psi = \pi/6$:
\begin{align}
\delta I^{(n)} = I(n\varOmega)\vert_{\psi=0} - I(n\varOmega)\vert_{\psi=\pi/6}.
\label{eq:diff_intensity}
\end{align}
Figure~\ref{fig:perturbative}(c) presents $\delta I^{(n)}$ on a logarithmic scale for $n=1$ to $9$.
For the first and third harmonics, we observe that
\begin{align}
\delta I^{(1)} \propto F_0^6 = o(F_0^2), \quad
\delta I^{(3)} \propto F_0^8 = o(F_0^6).
\end{align}
This suggests that a weak angle dependence, vanishing in the limit of $F_0 \to 0$, arises from higher-order perturbative processes.
On the other hand, for the fifth and higher harmonics, we see that
\begin{align}
\delta I^{(n)} \propto F_0^{2n},
\end{align}
which indicates that the angle dependence remains even in the limit of $F_0 \to 0$.
The above discussion and the numerical results shown in Fig.~\ref{fig:perturbative} explain why the nodelike structure observed for the fifth and seventh harmonics in Figs.~\ref{fig:lpdirection_j3} and \ref{fig:lpdirection_j3_resonant} is absent for the first and third harmonics; that is, when $F_0 = 1$, the first- and third-order responses are still near the perturbative regime, as tunneling excitation is negligible.

\bibliography{reference,reference_unpublished}

\end{document}